\documentclass[10pt,a4paper]{article}
\usepackage{jheppub_kim}
\usepackage{pdflscape}
\usepackage{amsmath}
\usepackage{amssymb}
\usepackage{dcolumn}
\usepackage{bm}
\usepackage{color}
\usepackage{epsfig}
\usepackage{amsfonts}
\usepackage{graphicx}
\usepackage{subfigure}
\usepackage{dcolumn}
\UseRawInputEncoding

\begin{document}
\title {Warm Inflation in $f(Q)$ gravity}
\author[a]{Tuhina Ghorui,}
\author[b]{Prabir Rudra,}
\author[c]{Farook Rahaman,}
\author[d,e,g]{Behnam Pourhassan}

\affiliation[a]{Department of Mathematics, Jadavpur University,
Kolkata-700 032, India.}

\affiliation[b] {Department of Mathematics, Asutosh College,
Kolkata-700 026, India.}

\affiliation[c] {Department of Mathematics, Jadavpur University,
Kolkata-700 032, India.}

\affiliation[d]{School of Physics, Damghan University, Damghan 3671645667, Iran.}
\affiliation[e]{Center for Theoretical Physics, Khazar University, 41 Mehseti Street, Baku, AZ1096, Azerbaijan.}
\affiliation[g]{Canadian Quantum Research Center 204-3002 32 Ave Vernon, BC V1T 2L7 Canada.}

\emailAdd{tuhinaghorui.math@gmail.com}
\emailAdd{prudra.math@gmail.com}
\emailAdd{farookrahaman@gmail.com}
\emailAdd{rahaman@associates.iucaa.in}
\emailAdd{b.pourhassan@du.ac.ir}

\abstract{
We investigate warm inflation in the framework of $f(Q)$ gravity within a Friedmann-Robertson-Walker spacetime. Unlike cold inflation, where the inflaton evolves in isolation, warm inflation features continuous interaction between the inflaton field and radiation throughout the inflationary epoch, facilitating energy transfer through dissipative processes and maintaining thermal equilibrium. In our novel approach, we employ $f(Q)$ dark energy as the driving mechanism for warm inflation, leveraging the geometric degrees of freedom associated with non-metricity as dynamical variables. We derive the field equations using slow-roll approximations and analyze two specific $f(Q)$ models: a power-law form $f(Q) = Q + mQ^n$ and a logarithmic form $f(Q) = mQ\ln(nQ)$. Our analysis focuses on the high-dissipative regime, where thermal fluctuations dominate over quantum fluctuations. We compute key inflationary observables, including the scalar spectral index $n_s$, tensor-to-scalar ratio $r$, and slow-roll parameters. Our results demonstrate that $f(Q)$ dark energy successfully drives warm inflation while satisfying essential physical conditions: initial dominance of $f(Q)$ energy density over radiation density initially, and thermal fluctuations exceeding quantum fluctuations ($T > H$). As inflation progresses, energy transfers from the geometric $f(Q)$ sector to radiation, eventually bringing both densities to comparable levels near inflation's end. Importantly, our computed values align well with current observational constraints from Planck and BICEP/Keck: $n_s = 0.965 \pm 0.004$ and $r < 0.036$. This validates the viability of warm inflation in $f(Q)$ gravity and establishes a unified geometric framework for understanding both early universe inflation and late-time cosmic acceleration.}


\maketitle

\section{Introduction}
Significant theoretical and observational cosmology has been conducted in recent years to understand the dynamics of the universe. The results of these investigations are in excellent agreement with the general relativity-based cosmological model known as the $\Lambda$CDM. Simultaneously, surveys looking at the cosmic microwave background (CMB) radiation have revealed important details about the beginnings and evolution of the universe \cite{GH, NA}. The conventional cosmological model still has several unclear concepts, such as the horizon and flatness is-sues. The concept of cosmological inflation has success-fully addressed some issues of standard cosmology. Infla-tion also provides a framework for understanding the generation and the evolution of the earliest seeds that gave rise to the universe's large-scale structure \cite{AAS, AHG, AAP, ADL, AL}.

Alan Guth pioneered the concept of inflation \cite{AHG} in 1981, which Sato \cite{S} later proposed independently. According to this concept, the first-order transition to true vacuum is used in de Sitter inflation. The universe would undergo a phase change if it cooled to temperatures 28 orders of magnitude lower than the critical temperature in its early history. In such scenarios, many issues such as flatness and horizon problems would vanish. A peri-od of exponential development would therefore produce a massive expansion factor, and when the latent heat is set free, the universe's entropy would be compounded by a large factor. In the framework of grand unified models of fundamental particle interactions, such a situation is entirely normal. The issue of monopole suppression in these models is similarly impacted by supercooling. The universe will thermalize during the reheating phase that follows inflation as the inflaton breaks down into light particles. It is possible to establish a connection between the parameters of inflation and reheating by considering the cosmic expansion history from the moment the meas-ured CMB scales escape the Hubble bounds during infla-tion to the moment they re-enter it later. A comprehens-ive review of cosmological inflation can be found in \cite{rev1}. Other notable works can be found in \cite{rev2, rev3, rev4, rev5, rev6, rev7, rev8}.

However, Guth and other authors \cite{prob1, prob2, prob3} who have since examined this issue have acknowledged that this scenario in the form proposed in \cite{AHG} has certain shortcom-ings. Therefore, this initial idea of inflation has been modified over the years to address these shortcomings. The modified version was first proposed by Linde \cite{L} and then by the authors of \cite{AS} in 1982. The concept uses the second-order transition of vacuum to explain the slow-roll inflation. Nevertheless, the idea also has drawbacks because it does not address the issue of spending too much time in a false vacuum, which can lead to massive inflation. Linde \cite{Li} studied chaotic inflation, which is a slow-roll inflation using initial disordered scalar fields \cite{shokri1}. According to this model, sections with substantial inflation produce an isotropic and homogeneous universe. Early and newly developed inflation models contend that the universe has consistently maintained thermal equilibrium, while chaotic inflation holds that this thermal equilibrium was not required. The problems with initial conditions are also resolved by the fact that disordered inflation matches with Planck density.

In the above discussion, we have seen the salient features of classical inflation or cold inflation where there is no role played by radiation. Theoretically, there is another form of cosmological inflation known as warm inflation \cite{warm1, ABP}. The dissipative effects are absent in cold inflation, meaning that the system's total energy is conserved in a conventional inflaton scalar field that is rolling down slowly. Meanwhile, warm inflation is a pro-cess where radiation production takes place during the in-flationary era and produces particles at relativistic scales. Because of de-Sitter-type expansion, particle density in conventional inflation exponentially decreases to zero following inflation. However, due to coupling with anoth-er field, the particle number does not become zero in the of warm inflation. The equation of conservation of the homogeneous scalar field contains an additional term involving a dissipative coefficient $\Gamma$, which maintains ra-diation production in the models of warm inflation. This term can be taken to be either a constant, a function de-pendent on temperature, scalar field, or a combination of the two. The second law of thermodynamics states that $\Gamma$ should be constrained as $\Gamma>0$ to couple with the radi-ation density. This means that throughout the inflationary phase, the radiation and inflaton field interact with each other. The dissipation term given by $\delta=\frac{\Gamma}{3H}$ is used to measure the importance of warm inflation ($H$ is the Hubble parameter). Strong dissipation $(\delta>> 1)$ and weak dissipation $(\delta<< 1)$ are the two extremes in warm inflation caused by the factor $\Gamma$. The weak dissipative regime is still dominated by the Hubble damping term, but the sluggish evolution of the inflaton field is controlled by the dissipative coefficient $\Gamma$ in the strong dissipative regime. Warm inflationary mechanism has been broadly studied in literature in different settings \cite{w1, w2, w3, w4, w5, w6, w7, w8, w9, w10, wam1, wam2, wam3, wam4, wam5, wam6, Saeed}.

The Type Ia Supernovae data \cite{RAF} clearly show that the universe is undergoing a second phase of accelerated expansion following the early inflation. The recent cosmic acceleration is attributed to the presence of a negative pressure component popularly known as dark energy (DE) \cite{de1}. In an alternative formulation, we modify the standard Einstein-Hilbert action producing the alternative theories of gravity \cite{md1}. Two fundamental formulations of gravity are found in the literature: the teleparallel ($R=0$, $\tau\neq 0$) and the curvature ($R\neq 0$, $\tau=0$) formulations, where $\tau$ is the scalar torsion and $R$ is the scalar curvature. Nevertheless, the non-metricity $Q$ disappears in both of these two formulations. The geometric representation of a vector's length variation in parallel transport is given by $Q$. There exists, however, a special type of non-metricity that admits fixed-length vectors. The mathematical foundations and physical implications of this special type of nonmetricity have been explored in \cite{new1, new2, new3, new4}. In a third comparable formalism of general relativity, it was previously believed that the elementary geometrical variable responsible for all sorts of gravitational interplay was a non-vanishing non-metricity $Q$. This theory has been named as the Symmetric Teleparallel gravity (STG) \cite{q1}. In this case, the Einstein pseudotensor represents the density of the energy-momentum tensor (EMT).  It should be noted that, when we geometrically represent this tensor, it ultimately becomes a real tensor. After much investigation, STG was expanded to $f(Q)$ gravity \cite{q2}, which is also referred to as non-metric gravity and coincident general relativity. The cosmological implications of $f(Q)$ gravity and its empirical limitations were studied in \cite{q3,q4}. In the past couple of decades, the STG framework has undergone a number of modifications \cite{q5, q6, q7, q8, q9}. By considering the non-minimal (NM) connection between the non-metricity $Q$ and the matter Lagrangian $L_m$, the authors of Ref. \cite{q9} have proposed extending the periphery of $f(Q)$ gravity. The NM connection between the geometry and matter sectors is expected to result in the non-preservation of the EMT and the addition of an extra force to the geodesic equation of motion. In \cite{q10}, the authors developed the $f(Q,T)$ model, which is another generalization of the STG. For example, see an extended study on inflation in $f(R)$, $f(T)$ and $f(Q)$ gravities \cite{shokri2}. In this case, the gravity Lagrangian is simply an arbitrary function of $Q$ and the trace $T$ of the EMT. The field equations were established, and the cosmic growth of the model was investigated.\\

The choice of $f(Q)$ gravity as the theoretical framework for warm inflation is motivated by several compelling theoretical and phenomenological considerations that go beyond its role as an alternative to General Relativity. First, the non-metricity scalar $Q$ naturally provides a geometric origin for the inflaton field, eliminating the need to introduce an ad hoc scalar field from particle physics. This geometric inflaton emerges directly from the spacetime structure itself, offering a more fundamental approach to early universe dynamics. From a thermodynamical perspective, $f(Q)$ gravity exhibits unique properties that are particularly well-suited for warm inflation scenarios. The non-metricity tensor $Q_{\alpha\mu\nu} = \nabla_\alpha g_{\mu\nu}$ captures information about the failure of the metric to be preserved under parallel transport, which can be interpreted as encoding microscopic degrees of freedom that naturally couple to thermal fluctuations. This geometric interpretation provides a physical basis for the dissipative interactions between the gravitational sector and radiation that are essential for warm inflation. It should be noted here that this is meaningful only within certain interpretation-al frameworks. Non-metricity does measure failure of metric preservation under parallel transport, but interpreting it as microscopic degrees of freedom coupled to thermal fluctuations is model-dependent and emergent, not a generic consequence of differential geometry. The idea that non-metricity encodes microscopic degrees of freedom comes from emergent gravity and spacetime thermodynamics, not from geometry alone. When the spacetime has microstructure, metric non-preservation reflects statistical fluctuations, which mirrors Brownian motion and hydrodynamics. In entropic gravity, non-metricity contributes to entropy production, modifies local Clausius relations, and encodes irreversible processes. This is explicit in non-equilibrium spacetime thermodynamics. Non-metricity naturally leads to non-conservation of length (dissipation), affine heat fluxes, and entropy generation. This matches with non-equilibrium thermodynamics and bulk viscosity-like terms. Hence, non-metricity is often interpreted as a dissipative geometric sector that brings in the thermal interpretation. In this connection, it is recommended that the reader consult some relevant literature \cite{q2, cpc1, cpc2, cpc3, cpc4, cpc5, cpc6, cpc7}. Moreover, $f(Q)$ gravity offers distinct advantages in addressing the fundamental challenges of inflationary cosmology. Unlike scalar field models, where the potential must be carefully fine-tuned to achieve slow-roll conditions, $f(Q)$ modific-ations can naturally produce the required flat effective potential through the functional form $f(Q)$ itself. The flexibility in choosing $f(Q)$ allows for a broader class of inflationary dynamics while maintaining theoretical consistency. The energy-momentum non-conservation inherent in $f(Q)$ gravity, expressed through the modified continuity equation $\nabla_\mu T^{\mu\nu} \neq 0$, provides a natural mechanism for energy transfer between the geometric sector and matter fields. This is particularly relevant for warm inflation, where continuous energy dissipation from the inflaton to radiation is required throughout the inflationary epoch. In contrast to scalar field theories, where such dissipation must be imposed through additional coupling terms, $f(Q)$ gravity incorporates this energy exchange as a fundamental feature of the modified gravitational dynamics. Furthermore, recent studies have shown that $f(Q)$ theories can naturally accommodate both early-time inflation and late-time cosmic acceleration within a unified framework \cite{Lazkoz2019, Solanki2021}. This dual capability suggests that $f(Q)$ gravity may provide a more economical description of cosmic evolution, potentially resolving the coincidence problem that plagues multi-component dark sector models. The non-metricity approach also offers advantages in terms of observational predictions. The additional degrees of freedom in $f(Q)$ gravity can lead to distinctive signatures in the cosmic microwave background, particularly in the tensor-to-scalar ratio and spectral indices, that differ from both standard inflation and other modified gravity scenarios. These signatures provide testable predictions that can distinguish $f(Q)$ warm inflation from alternative models. Additionally, $f(Q)$ gravity maintains the general covariance of General Relativity while avoiding some of the pathologies associated with higher-derivative theories, such as $f(R)$ gravity. The second-order field equations ensure the absence of Ostrogradsky instabilities, making $f(Q)$ gravity a theoretically robust framework for cosmological applications. The realm of truth of the statement is constrained, and it requires some important qualifications. The second-order nature of the field equations in $f(Q)$ gravity guarantees the absence of Ostrogradsky instabilities associated with higher derivatives, making it a structurally robust framework. However, this is only true in the strict Ostrogradsky sense. The second-order field equations do not guarantee full stabil-ity \cite{ghost1, ghost2}. $f(Q)$ generically propagates one extra scalar mode, analogous to the scalar in $f(R)$, but without higher derivatives. This scalar can suffer from ghost, instability, gradient, instability, and tachyonic instabilities. These are not Ostrogradsky ghosts, but are still physical instabilities. Thus, the full theoretical viability requires additional stability condi-tions on the propagating scalar mode and cosmological perturbations. With proper choices of $f(Q)$, the theory is robust and competitive for cosmology.

It is known that DE is equivalent to modified gravity. We know that the best candidate to drive the early-time cosmic inflation is a scalar field. However, it has been shown in the literature that different DE models, such as holographic DE (HDE) and Chaplygin gas, etc., can act as suitable replacements for a scalar field in an inflationary scenario. As we know that DE is equivalent to a modific-ation to the standard gravity in explaining late-time accel-eration, it makes sense to explore the inflationary dynam-ics of the universe in modified gravity. Motivated by this, we attempt to investigate warm inflation in $f(Q)$ gravity. Although there are some studies have explored cosmological inflation has been explored in the framework of $f(Q)$ gravity, there have been very few attempts to study a warm inflationary mechanism in the background of $f(Q)$ gravity. Two notable works in this direction can be found in \cite{wife1, wife2}. In \cite{wife1}, the authors have studied a warm inflationary scenario triggered by entropies of some recent DE models within $f(Q)$ grav-ity. To accomplish this task, they introduced the Tsallis, Renyi, and Barrow holographic dark energy (HDE) entropies into the standard Friedmann equations. Our plan is different compared to this work in the sense that we will not include any DE model in our work. Because DE is considered to be equivalent to modified gravity, we will use the DE compon-ent arising from the $f(Q)$ gravity as the driving component of the warm inflation. Thus, the entire scheme of our work is different from that in \cite{wife1}. Moreover, our analys-is will help us to understand the effectiveness of the dark component coming purely from $f(Q)$ gravity in driving warm inflation. In \cite{wife2},  the authors have studied the perturba-tion spectra of warm inflation in $f(Q,T)$ gravity, which is a further modification of $f(Q)$ gravity. This work used a different gravity theory than our plan of work. We admit that $f(Q, T)$ gravity is just a derivative theory of the $f(Q)$ the-ory, and shares the same origin. Thus they may have some like properties, but eventually, $f(Q,T)$ gravity will vary from $f(Q)$, because of the contribution  from the matter component. Moreover, in \cite{wife2}, the authors have studied the perturbation spectra of warm inflation, whereas our plan of work does not include any perturbation spectra analysis. Our work plan is limited to cosmological inflationary parameters and aims at to study the effectiveness of pure $f(Q)$ DE in driving the warm inflation. Thus, in this sense, our work is quite different and original. The remainder of this article is arranged as follows: In section 2, we discuss warm inflationary mechanism in $f(Q)$ gravity. In section 3, we explore how $f(Q)$ DE can drive warm in-flation. In section 4, we discuss the results obtained in the study, and finally the paper ends with a conclusion in section 5.

\section{An overview of Warm inflation in $f(Q)$ gravity}
We will start with the following action of the $f(Q)$ gravity:
\begin{equation}\label{action}
S=\int\left[-\frac{1}{2\kappa^{2}}f(Q)+\mathcal{L}_{m}\right]\sqrt{-g}d^{4}x
\end{equation}
where $\mathcal{L}_{m}$ is the matter Lagrangian, $f(Q)$ is an
arbitrary function of the non-metricity scalar $Q$, and $g$ is the
determinant of the metric tensor $g_{\mu\nu}$. The
non-metricity scalar is defined as
\begin{equation}\label{nm1}
Q=-\frac{1}{4}Q_{\alpha\beta\gamma}Q^{\alpha\beta\gamma}+\frac{1}{2}Q_{\alpha\beta\gamma}Q^{\gamma\beta\alpha}+\frac{1}{4}Q_{\alpha}Q^{\alpha}-\frac{1}{2}Q_{\alpha}\Tilde{Q}^{\alpha},
\end{equation}
where
\begin{equation}\label{nm2}
Q_{\alpha}\equiv Q_{\alpha\mu}^{\mu},
\end{equation}
\begin{equation}\label{nm3}
\Tilde{Q}^{\alpha}\equiv Q_{\mu}^{\mu\alpha}
\end{equation}
and the non-metricity tensor can be given by,
\begin{equation}\label{nm4}
Q_{\alpha\mu\nu}\equiv\nabla_{\alpha}g_{\mu\nu}.
\end{equation}
If we take $f(Q)=Q$, then we get the Symmetric Teleparallel
Equivalent of General Relativity (STEGR). Now from the equations
(\ref{action}),(\ref{nm1}), (\ref{nm2}), (\ref{nm3}), (\ref{nm4}),
the field equations are generated as
\begin{eqnarray*}\label{field1}
\frac{2}{\sqrt{-g}}\nabla_{\alpha}\Big\{\sqrt{-g}g_{\beta\nu}f_{Q}\Big[-\frac{1}{2}L^{\alpha\mu\beta}+\frac{1}{4}g^{\mu\beta}(Q^{\alpha}-\Tilde{Q}^{\alpha})-\frac{1}{8}(g^{\alpha\mu}Q^{\beta}+g^{\alpha\beta}Q^{\mu})\Big]\Big\}
\end{eqnarray*}
\begin{equation}
+f_{Q}\Big[-\frac{1}{2}L^{\mu\alpha\beta}-\frac{1}{8}(g^{\mu\alpha}Q^{\beta}+g^{\mu\beta}Q^{\alpha})+\frac{1}{4}g^{\alpha\beta}(Q^{\mu}-\Tilde{Q}^{\mu})\Big]Q_{\nu\alpha\beta}+\frac{1}{2}\delta^{\mu}_{\nu}f=\kappa^{2}T^{\mu}_{\nu},
\end{equation}
where $f_{Q}\equiv\frac{\partial f}{\partial Q}$. Here, the
deformation tensor is given by
\begin{equation}\label{deform1}
L^{\alpha}_{\mu\nu}=\frac{1}{2}Q^{\alpha}_{\mu\nu}-Q_{\mu
\nu}^{\alpha}
\end{equation}
and the matter energy-momentum tensor is
\begin{equation}\label{emt1}
T_{\mu\nu}=-\frac{2}{\sqrt{-g}}\frac{\delta\left(\sqrt{-g}\mathcal{L}_{m}\right)}{\delta g^{\mu\nu}}
\end{equation}
We consider a spatially flat, homogeneous, and isotropic universe modelled by the Friedmann-Lemaitre-Robertson-Walker (FLRW) metric:
\begin{equation}
ds^{2}=-dt^{2}+a^{2}(t)\left(dx^{2}+dy^{2}+dz^{2}\right)
\end{equation}
where $a(t)$ is the cosmological scale factor that accounts for the expansion of the universe. Two modified Fried-mann equations of $f(Q)$ gravity that serve as the primary dynamical equations are given by
\begin{equation} \label{f}
H^{2}=\frac{1}{M_{p}^{2}}\left(\rho_{r}+\rho_{Q}\right)
\end{equation}
and
\begin{equation} \label{h}
\dot{H} =-\frac{1}{2{M_{p}^{2}}}(\rho_{Q}+p_{Q}+\rho_{r}+p_{r})
\end{equation}
Where $H=\frac{\dot{a}}{a}$ is the Hubble rate of expansion and $M_{p}=\frac{1}{\sqrt{8 \pi G}}\approx 1$ is the reduced plank mass. Moreover, the subscripts $r$ and $Q$ represent the contributions from radiation and modified gravity, respectively. The energy density and pressure contributions of $f(Q)$ gravity are respectively given by
\begin{equation}\label{denq}
\rho_{Q}=M_{p}^{2}\left(3 H^{2}(1+2 f_{Q})-\frac{f}{2}\right)
\end{equation}
and
\begin{equation}\label{pressureq}
p_{Q}=-M_{p}^{2}\left(2\dot{H}(1+f_{Q})-\frac{f}{2}+3H^{2}(1+2 f_{Q}+8 f_{QQ}\dot{H})\right)
\end{equation}
where $f=f(Q)$, $f_{Q}=\frac{\partial f}{\partial Q}$, and $f_{QQ}=\frac{\partial^{2}f}{\partial Q^{2}}$. Moreover, $\dot{H}$ is the time derivative of the Hubble parameter.\\

The identification of $f(Q)$ DE with the inflaton field represents a paradigm shift from traditional scalar field inflation models and requires careful physical justifica-tion. In our approach, the geometric $f(Q)$ sector plays the dual role of both the source of inflationary dynamics and the entity that drives the current cosmic ac-celeration, providing a unified description of the universe's two accelerated expansion phases. The physical basis for this identification rests on several key observations. First, the effective energy density and pressure of the $f(Q)$ sector, given by equations (\ref{denq}) and (\ref{pressureq}), naturally exhibit the negative pressure behavior ($p_Q < 0$) required for accelerated expansion. The time evolution of these quantities is governed by the modified Friedmann equations, which incorporate the geometric degrees of freedom associated with non-metricity as dynamical variables. Crucially, the $f(Q)$ energy density $\rho_Q$ behaves as an effective perfect fluid with an equation of state that can vary dynamically depending on the functional form of $f(Q)$ and the cosmological epoch. During inflation, the choice of appropriate $f(Q)$ functions can ensure that $w_Q = p_Q/\rho_Q \approx -1$, mimicking the behavior of a slow-roll scalar field. This is achieved through the slow-roll conditions expressed in terms of the non-metricity scalar rather than a traditional potential. The advantage of this geometric approach becomes evident when considering the origin of the inflaton's energy. In scalar field models, one must postulate the existence of a fundamental scalar field with a carefully crafted potential $V(\phi)$ to achieve inflation. In contrast, $f(Q)$ gravity derives the inflationary energy directly from the geometric structure of spacetime itself. The function $f(Q)$ modifies the gravitational dynamics in a way that the geometric sector naturally acquires the energy-momentum characteristics necessary for inflation. From a field-theoretic perspective, the $f(Q)$ approach can be understood as an effective field theory where the non-metricity degrees of freedom are promoted to dynamical variables. The action principle with $f(Q)$ terms generates field equations that are mathematically equivalent to having an additional stress-energy tensor $T_{\mu\nu}^{(Q)}$ with the components,\\
\begin{eqnarray}
T_{00}^{(Q)} &=& \rho_Q = M_p^2 \left[ 3H^2(1 + 2f_Q) - \frac{f(Q)}{2} \right], \\
T_{ii}^{(Q)} &=& p_Q = -M_p^2 \left[ 2\dot{H}(1 + f_Q) - \frac{f(Q)}{2} + 3H^2(1 + 2f_Q + 8f_{QQ}\dot{H}) \right].
\end{eqnarray}
This effective stress-energy tensor arises purely from the geometric modification and does not require the introduction of additional matter fields. The evolution equations for $\rho_Q$ and $p_Q$ emerge naturally from the modified Einstein equations, providing a self-consistent description of the inflationary dynamics.

Inflation is not a standalone component of the model; rather, it interacts with other fields in any particle physics perception of the inflationary backdrop. As a result of these interactions, a tiny percentage of the inflation vacuum energy may be converted into other forms of energy, which could cause the inflation energy to dissipate into different degrees of freedom. The two-stage warm inflation mechanism involves dissipation, leading to particle production. We can model the contribution of those relativistic particles as radiation when they thermalize quickly enough, (for example, in less than a Hubble time) in an expanding universe. The energy density for radi-ation can be given by
\begin{equation} \label{j}
\rho_{r} \simeq \frac{\pi^{2}}{30}g_{*}T^{4}
\end{equation}
where $g_{*}$ is the effective number of light degrees of freedom, and $T$ is the thermal bath's temperature. The in-flaton field $\phi$ responsible for inflation can be given by the following equation:
\begin{equation} \label{a}
\ddot{\phi}+(3H+\Gamma)\dot{\phi}+V_{\phi}=0
\end{equation}
$\Gamma$ is the dissipative coefficient, which could be either constant, depend on the scalar field or temperature, or depend on both scalar field $\phi$ and temperature $T$.

The dissipative coefficient $\Gamma$ in equation (\ref{a}) represents the crucial link between the geometric $f(Q)$ inflaton and the thermal radiation bath. Unlike phenomenological ap-proaches that treat $\Gamma$ as an arbitrary function, a complete theory requires understanding its microscopic origin from fundamental particle physics interactions.\\
In the context of $f(Q)$ gravity, the non-metricity tensor $Q_{\alpha\mu\nu}$ can be decomposed into its irreducible components, including vectorial, tensorial, and scalar parts. These geometric degrees of freedom couple to matter fields through the non-minimal coupling terms that naturally arise in $f(Q)$ theories. The effective action can be written as,
\begin{equation}
S_{\text{eff}} = \int d^4x \sqrt{-g} \left[ -\frac{1}{2\kappa^2} f(Q) + \mathcal{L}_m + \mathcal{L}_{\text{int}} \right]
\end{equation}
where $\mathcal{L}_{\text{int}}$ represents the interaction Lagrangian between the geometric $f(Q)$ sector and matter fields. The interaction Lagrangian can be parametrized as,
\begin{equation}
\mathcal{L}_{\text{int}} = -\frac{\lambda}{4} Q_{\alpha\mu\nu} J^{\alpha\mu\nu} - \frac{g}{2} Q \phi^2 - \frac{h}{3!} Q^{1/2} \phi^3 + \text{higher order terms}
\end{equation}
where $J^{\alpha\mu\nu}$ is the non-metricity current, $\phi$ represents light scalar fields (such as Standard Model fields), and $\lambda$, $g$, and $h$ are coupling constants that depend on the specific particle content and symmetries of the theory.\\
Following the approach developed for warm inflation in supersymmetric theories \cite{Berera1995,Bastero-Gil2011}, we can derive the dissipative coefficient from first principles. The key insight is that the $f(Q)$ geometric fluctuations decay into light degrees of freedom through the interaction terms in $\mathcal{L}_{\text{int}}$.\\
Consider a simple model where the $f(Q)$ sector couples to $N_f$ light fermion fields through,
\begin{equation}
\mathcal{L}_{\text{int}} = -g_f \frac{Q^{1/2}}{M_p} \bar{\psi}_i \psi_i
\end{equation}
where $g_f$ is a dimensionless coupling constant and $M_p$ is the Planck mass. Using thermal field theory techniques, the one-loop contribution to the dissipative coefficient is,
\begin{equation}
\Gamma_1 = \frac{g_f^2 N_f T^3}{32\pi^3 M_p^2} \int_0^\infty  \, x^2 \frac{e^x}{(e^x - 1)^2} dx = \frac{g_f^2 N_f T^3}{12 M_p^2}
\end{equation}
For coupling to bosonic fields (such as gauge bosons or scalars), the calculation yields,
\begin{equation}
\Gamma_{\text{bos}} = \frac{g_b^2 N_b T^3}{8 M_p^2} \left[ 1 + \frac{T^2}{12 M_p^2} f''(Q) \right]
\end{equation}
where $N_b$ is the number of bosonic degrees of freedom, and $g_b$ is the corresponding coupling.\\
The total dissipative coefficient combines contributions from all light degrees of freedom:
\begin{equation}
\Gamma_{\text{total}} = \Gamma_1 + \Gamma_{\text{bos}} + \Gamma_{\text{gauge}} + \Gamma_{\text{higher-loop}}.
\end{equation}
In the high-temperature regime ($T \gg M_{\text{EW}}$, where $M_{\text{EW}}$ is the electroweak scale), all Standard Model degrees of freedom contribute, giving approximately,
\begin{equation}
\Gamma \approx \frac{C_{\text{SM}} g_{\text{eff}}^2 T^3}{M_p^2}
\end{equation}
where $C_{\text{SM}} \approx 106.75$ accounts for the Standard Model degrees of freedom, and $g_{\text{eff}}^2$ is an effective coupling that depends on the specific $f(Q)$ model.

Here, we will consider that inflation is driven by $f(Q)$ modification to standard gravity. The inflation energy density will therefore be given by equation(\ref{denq}). The equation for the evolution of the inflation energy density $\rho_{Q}$ is given by the equation (\ref{a}):
\begin{equation} \label{b}
\dot{\rho_{Q}}+3H(\rho_{Q}+p_{Q})=-\Gamma(\rho_{Q}+p_{Q})
\end{equation}
Therefore, to maintain energy conservation, the radiation fluid $\rho_{r}$ must gain the energy lost by the inflation field, with the right hand side of equation (\ref{b}) serving as the source term. Thus, we have
\begin{equation}
\dot{\rho_{r}}+3H(\rho_{r}+p_{r})=\Gamma(\rho_{Q}+p_{Q})
\end{equation}
The dissipative character essential for warm inflation is incorporated through the coupling between the $f(Q)$ sector and radiation, as expressed in equation (\ref{b}). This coupling represents the energy transfer from the geometric degrees of freedom to the thermal radiation bath. Physically, this can be interpreted as the decay of non-metricity fluctuations into relativistic particles, similar to how inflaton fluctuations decay in traditional warm inflation models. An important advantage of the $f(Q)$ approach is that it naturally avoids the $\eta$-problem that plagues many scalar field inflation models. The slow-roll parameters in $f(Q)$ inflation are determined by the derivatives of the function $f(Q)$ rather than by the fine-tuning of a scalar potential. This provides greater flexibility in constructing viable inflationary models without requiring extreme fine-tuning of parameters. Furthermore, the $f(Q)$ DE approach offers a natural explanation for the ob-served similarity between the vacuum energy scales of early inflation ($\sim 10^{16}$ GeV) and late-time DE ($\sim 10^{-3}$ eV) when viewed through the lens of the running of the func-tion $f(Q)$ with the cosmic evolution. Different regimes of the non-metricity scalar $Q$ can correspond to different cosmological epochs, providing a unified framework for understanding both phases of cosmic acceleration. The mathematical consistency of treating $f(Q)$ DE as the in-flaton is ensured by the second-order nature of the field equations in $f(Q)$ gravity, which avoids the ghost in-stabilities that can arise in higher-derivative theories. The dynamics are well-defined and lead to physically reason-able predictions for observational quantities, such as the scalar and tensor power spectra.

In the case of classical cold inflation, there are slow-roll estimates, which are typically defined as the slow-roll parameters. The first slow-roll parameter is defined as
\begin{equation} \label{m}
\epsilon_{1}=-\frac{\dot{H}}{H^{2}}=-\frac{d\ln{H}}{dN}
\end{equation}
where $dN=Hdt$. Now, using a generalized relation, the next slow-roll parameters are defined as
\begin{equation} \label{m1}
\epsilon_{n+1}=-\frac{\dot{\epsilon_{n}}}{H\epsilon_{n}}
\end{equation}
Furthermore, there exists an alternative form of the slow-roll parameter in the case of warm inflation, de-noted by
\begin{equation} \label{n}
\beta=\frac{\dot{\Gamma}}{H\Gamma}
\end{equation}
This parameter characterizes how the dissipation coefficient changed for inflation. The Hubble rate of expansion can be found in terms of the scalar field value, and vice versa, using the slow-roll approximation. According to the Hubble parameter, the total logarithmic expansion, or the number of "e-folds," can be given with respect to the scalar field value during inflation. Now, we can define the number of e-foldings $N$ between two possible cosmological times $t_1$ and $t_2$, where $t_1$ is the time of horizon crossing and $t_2$ corresponds to the end of inflation: 
\begin{equation} \label{t}
N=\int_{t_1}^{t_2}H dt
\end{equation}
The parameter $N$ measures the expansion of the universe during inflation. This relationship could be used to differentiate a parameter's value from its beginning value at inflation to its end value.

When there are sufficient particles during the inflationary age, warm inflation takes place. At a temperature $T$, we will suppose that there are sufficient particle interactions to generate a thermal radiation gas. Both thermal and quantum fluctuations are present in the warm inflation scenario, with the thermal fluctuations prevailing as long as $T$ is higher than $H$ \cite{AB, LIA, IC, RL}. The scalar perturbations amplitude \cite{MA, AJ} is determined by
\begin{equation}
P_{s}=\frac{H^{2}}{8\pi^{2}M_{p}^{2}\epsilon_{1}}\left(1+2n_{BE}+\frac{2\sqrt{3}\pi \delta}{\sqrt{3+4\pi \delta}}\frac{T}{H}\right)G(\delta)
\end{equation}
The radiation bath's presence generates the statistical distribution of the inflaton, which is denoted by $n_{BE}= \left({e^{\frac{H}{T_{in}}}-1}\right)^{-1}$, known as the Bose-Einstein distribution, and $T_{in}$ is the inflaton fluctuation. The function $G(\delta)$ is given by
\begin{equation}
G(\delta)=1+0.0185~ \delta^{2.315}+0.335~ \delta^{1.364}
\end{equation}
where $G(\delta)$ is the growth of inflation fluctuations brought on by the coupling to radiation and can be determined numerically \cite{MA, gq1}. In an inflationary setting, two commonly utilized parameters are the tensor-to-scalar ratio ($r$) and the scalar spectral index ($n_s$). The scalar spectral index can be defined by
\begin{equation}
n_{s}=1+\frac{dln(P_{s})}{d\ln{k}}
\end{equation}
where $k=aH$ or $\ln{k}=N+\ln{H}$. Scalar spectral index measures the scale dependence ("tilt") of scalar perturbations. When $\delta\rightarrow0$ and $T\rightarrow 0$, the spectral index \cite{AD, LMH, bauman} for supercooled inflation becomes $n_{s}=1-6\epsilon+2\eta$. From the observational data, the measure of the spectral index is obtained as
$n_{s}=0.9649 \pm 0.0042$ \cite{y}, which is in the proximity of $1$.

On super-Hubble scales, tensor modes are locked in both during and after inflation. Thus, the first vacuum fluctuations on sub-Hubble scales are associated with the late-time power spectrum for tensor modes at Hubble departure during inflation. The amplitude of tensor perturbation \cite{MA,AJ} is expressed as
\begin{equation}
P_{t}=\frac{2H^{2}}{\pi^{2} M_{p}^{2}}
\end{equation}
Using this perturbation parameter, the tensor-to-scalar ratio $r$ is defined by
\begin{equation} \label{w}
r=\frac{P_{t}}{P_{s}}
\end{equation}
This parameter compares the amplitude of tensor modes to scalar modes. Using the most significant observations, the tensor-to-scalar ratio $r$ and the spectral index $n_{s}$ of the primordial scalar curvature perturbations can also be written in terms of the slow-roll parameters as,
$r=16\epsilon_{1}, n_{s}=1-4\epsilon_{1}-2\epsilon_{2} $ or as $r \cong 16\epsilon, n_{s}\cong 1-6\epsilon+2\eta $, where $\epsilon=\epsilon_{1}$   and $\eta=\epsilon_{1}-\epsilon_{2}$.
The parameter still lacks precise data, and the most recent observational data only suggests an upper limit on $r$ of $r<0.064$ \cite{y}. From the re-cent Planck 2018 and BICEP/Keck experiments, we see that tighter observational constraints have been imposed on the inflationary parameters. These observations imposed the following constraints on the parameters: $n_s=0.965\pm 0.004$, ~$r<0.036$.

\section{$f(Q)$ dark energy for Warm Inflation}
In this section, we will use the contribution from $f(Q)$ gravity as the source of warm inflation. The value and the nature of the inflationary parameters depend entirely on the inflationary potential, so the form of the potential determines stability. Here, we consider that the inflationary potential is supplied by the $f(Q)$ modified gravity field. It is widely seen in the literature that the exotic DE models can act as the driver of warm inflation \cite{rev8, w9}. Since it is known that, physically, DE is equivalent to modified gravity, there is good motivation to use the contri-bution from a modified gravity like $f(Q)$ gravity as the source of warm inflation. The inflation energy density will be the same as the energy contribution of  $f(Q)$ grav-ity. Using (\ref{denq}) in the Friedmann equation (\ref{f}), we get
\begin{equation}\label{c}
H^{2}=\frac{1}{3M_{p}^{2}}\left[M_{p}^{2}\left(3H^{2}\left(1+2f_{Q}\right)-\frac{f(Q)}{2}\right)+\rho_{r}\right].
\end{equation}
The quasi-stable nature of radiation production during the inflationary era, i.e., $\rho_{r}<<4H\rho_{r}$, is given by
\begin{equation} \label{i}
    4H\rho_{r}=\Gamma(\rho_{Q}+p_{Q})
\end{equation}
Using (\ref{i}) in  the 2nd Friedmann equation (\ref{h}), we get the radiation energy density as
\begin{equation}\label{d}
\rho_{r}=-\frac{3M_{p}^{2}}{2}\frac{\delta}{1+\delta}\dot{H}
\end{equation}
where warm inflation's significance is typically measured by
$\delta=\frac{\Gamma}{3H}$. When equation (\ref{d}) is substituted into the Friedmann equation (\ref{c}), the Hubble parameter's time derivative is obtained as
\begin{equation} \label{l}
\dot{H}=-2\left[\frac{f(Q)}{Q}-2f_{Q}\right]\frac{1+\delta}{\delta}H^{2}
\end{equation}
The radiation energy density can be rewritten in terms of the function of $Q$ and the Hubble parameter as
\begin{equation}\label{k}
\rho_{r}=3M_{p}^{2}\left[\frac{f(Q)}{Q}-2f_{Q}\right]H^{2}
\end{equation}
If we consider thermalization, the energy density of the radiation field could be expressed as follows: 
\begin{equation} \label{o}
\rho_{r}=C_{r} T^{4}
\end{equation}
where $C_{r}\simeq 70$  is the Stephen-Boltzmann constant\cite{GV}.

Comparing equations (\ref{k}) and (\ref{o}), the temperature of the thermal bath can be expressed as
\begin{equation}\label{tempo}
T=\left[\frac{3M_{p}^{2}}{C_{r}}\left(\frac{f(Q)}{Q}-2f_{Q}\right)H^{2}\right]^{\frac{1}{4}}
\end{equation}
Using equation$(\ref{l})$ in the definition of $\epsilon_{1}$ in equa-tion.$(\ref{m})$, the first slow roll parameter can be expressed in the form of a function of $Q$ as
\begin{equation} \label{p}
\epsilon_{1}=2\left[\frac{f(Q)}{Q}-2f_{Q}\right]\frac{1+\delta}{\delta}
\end{equation}
Using $(\ref{l})$, $(\ref{k})$, $(\ref{n})$, and $(\ref{m})$, the 2nd slow roll para-meter can be expressed as
\begin{equation}
\epsilon_{2}=\epsilon_{1}\left[1-Q\left(\frac{f_{Q}+2Qf_{QQ}}{2Qf_{Q}-f(Q)}\right)\right]-\frac{1}{1+\delta}\left(\beta-\epsilon_{1}\right)
\end{equation}

As warm inflation develops, dissipative effects play a significant role.  These effects arise from the scalar field decaying into a thermal bath due to friction.  In the setting of supersymmetry, the dissipative coefficient $\Gamma$ has been evaluated from the fundamentals in \cite{w1, ar15}.  However, as demonstrated in \cite{kamali}, there are alternative methods of constructing warm inflation that do not rely on supersymmetry. Although the co-efficient may be seen as constant, it can also be viewed more broadly as a function of temperature $T$. The power law form of the temperature can be considered as,
\begin{equation}\label{3.10}
\Gamma=C_{T}T^{\gamma}
\end{equation}
Substituting equation (\ref{tempo}) into the above equation, we get
\begin{equation}
\Gamma=C_{T}\left[\frac{3M_{p}^{2}}{C_{r}}\left(\frac{f(Q)}{Q}-2f_{Q}\right)H^{2}\right]^{\frac{\gamma}{4}}
\end{equation}
Using equation (\ref{n}) and (\ref{l}), the parameter $\beta$ can be expressed as
\begin{equation}
\beta=-2\left(\frac{f(Q)}{Q}-2f_{Q}\right)\frac{1+\delta}{\delta}\frac{H\Gamma_{,H} }{\Gamma}
\end{equation}
In warm inflation, the slow-roll condition \cite{GV} can be indicated by $\epsilon_{1} \textless \textless 1+\delta $, $\epsilon_{n} \textless \textless 1+\delta $, $\beta \textless \textless 1+\delta$. For the power-law temperature dependence used in equation (\ref{3.10}), the microscopic derivation provides,
\begin{equation}
\gamma = \begin{cases}
1 & \text{(one-loop fermion dominated)} \\
3 & \text{(one-loop boson dominated)} \\
1-3 & \text{(mixed contributions)} \\
> 3 & \text{(higher-loop or non-perturbative effects)}
\end{cases}
\end{equation}

\subsection{Warm Inflation in high dissipative regime}
In this section, we will assume that inflation takes place in a high dissipative regime (HDR), i.e., $\delta \textgreater\textgreater 1$. Here, we will explore two toy models of $f(Q)$ gravity as special cases to demonstrate the warm inflationary setting developed in the previous section for $\delta \textgreater\textgreater 1$.

By applying this condition to the equation (\ref{p}), the first slow-roll parameter can be written as
\begin{equation}
\epsilon_{1}=-\frac{\dot{H}}{H^{2}}=2\left(\frac{f(Q)}{Q}-2f_{Q}\right)
\end{equation}
When the generalized expression gives the slow-roll parameters, the second slow-roll parameter can be furnished as
\begin{equation}
\epsilon_{2}=-\frac{\dot{\epsilon_{1}}}{H\epsilon_{1}}=\epsilon_{1}\left[1-Q\left(\frac{f_{Q}+2Qf_{QQ}}{2Qf_{Q}-f(Q)}\right)\right]
\end{equation}
After applying the condition of HDR, we get the Hubble parameter's time derivative, which can be rewritten as
\begin{equation} \label{s}
\dot{H}=-2\left(\frac{f(Q)}{Q}-2f_{Q}\right)H^{2}
\end{equation}

\subsubsection{Toy model-1}
For our analysis, we take into account a functional form \cite{SS} of $f(Q)$ that which consists of a linear term and a non-linear term of the non-metricity scalar $Q$, which is known as the power-law model. We present the model below:
\begin{equation} \label{u1}
f(Q)=Q+mQ^{n},
\end{equation}
where $m$ and $n$ are constants. The linear term $Q$ represents the classical Einstein-Hilbert action in the teleparallel formulation, while the $mQ^n$ term encodes quantum corrections. The power-law form naturally explains the hierarchy between early inflation and late-time acceleration. During inflation, when $Q \sim H^2 \sim (10^{13} \text{ GeV})^2$, the $Q^n$ term dominates for $n > 1$, driving inflation. In the current epoch, when $Q \sim H_0^2 \sim (10^{-33} \text{ eV})^2$, the linear term dominates, leading to the observed small cosmological constant. The coupling strength between the $f(Q)$ sector and Standard Model fields depends on the specific functional form of $f(Q)$. For the power-law model (\ref{u1}),
\begin{equation}
g_{\text{eff}}^2 \approx \frac{|m| n (n-1)}{M_p^{2(n-2)}} H^{2(n-2)}.
\end{equation}
This provides a direct connection between the phenomenological parameters $m$ and $n$ in the $f(Q)$ function and the underlying particle physics coupling. The dissipative coefficient becomes,
\begin{equation}
\Gamma = C_T \frac{|m| n (n-1)}{M_p^{2n-2}} H^{2(n-2)} T^{\gamma}
\end{equation}
where $C_T$ is a numerical factor depending on the particle content.
Substituting equations (\ref{u1}) and (\ref{t}) into equa-tion (\ref{s}), we obtain the Hubble parameter with respect to the e-folding number $N$ as 
\begin{equation} 
H^{2(n-1)}(N)=\frac{e^{4(n-1)N}}{m(1-2n)6^{(n-1)}e^{4(n-1)N}+c_{1}}
\end{equation}
where $c_{1}$ is a constant. Because both the slow-roll para-meters depend on the function of $Q$ expressed as a Hubble parameter, both parameters are written in terms of the e-folding number as
\begin{equation}
\epsilon_{1}(N)=-\frac{2c_{1}}{m(1-2n)6^{n-1}e^{4(n-1)N}+c_{1}}
\end{equation}
and
\begin{equation}
\epsilon_{2}(N)=-\frac{4m(1-2n)(1-n)6^{n-1}e^{4(n-1)N}}{m(1-2n)6^{n-1}e^{4(n-1)N}+c_{1}}
\end{equation}
The warm inflation parameter $\beta$ can be written in terms of the number of e-folds as
\begin{equation}
\beta=-\gamma\frac{m(1-2n)(n-1)6^{n-1}e^{4(n-1)N}-c_{1}}{m(1-2n)6^{n-1}e^{4(n-1)N}+c_{1}}
\end{equation}
For the HDR, the dissipative parameter $\delta>> 1$. Thus, we have $G(\delta) \approx 0.00185~ \delta^{2.315}$ \cite{w9}. The scalar spectral in-dex is obtained in terms of slow-roll parameters as
\begin{equation}\label{nss1}
n_s=1+1.815\epsilon_{1}-\epsilon_{2}+3.815\beta
\end{equation}
Using equations (\ref{w}), the tensor-to-scalar ratio is giv-en by
\begin{equation}\label{nss2}
r=16 \epsilon_{1} \left(\frac{T}{H}0.00185~ \delta^{2.815}\right)^{-1}
\end{equation}




\begin{figure}[hbt!]
\begin{center}
\includegraphics[height=2.5in]{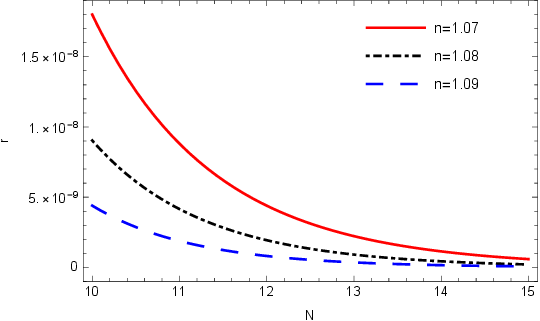}
\caption{The figure shows the tensor-to-scalar ratio $r$ with respect to the e-folding number $N$ for different values of the model para-meter $n$ for model-$1$. Other parameters are given by $m=4.5, c1=1, c_{r}=0.25$ .}
\label{figscale1}
\end{center}
\end{figure}

\begin{figure}[hbt!]
\begin{center}
\includegraphics[height=2.5in]{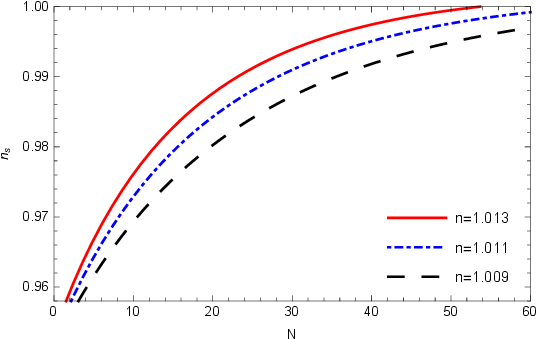}
\caption{Plot of the scaler spectral index $n_{s}$ against $N$ for different values of model parameter $n$ for model-1.  Other parameters are given by $c1 = 1,m = 4.5, \gamma = 1$ .}
\label{figscale2}
\end{center}
\end{figure}

\begin{table}
\caption{The table shows numerical values of the scalar spectral index and tensor-to-scalar ratio for different values of $N$ for model-1. The other parameters are taken as  $c_1=1$, $m=0.75$, $\gamma=0.99$.}

    \label{tab:my_label}
    \vspace{5mm}
    \centering
    \begin{tabular}{||l|c|c|r||}
    \hline
        N & n & $n_{s}$ &r\\ [1ex]
   
   \hline \hline
   50  & 1.007 & 0.933124 & $3.99534\times10^{-8}$\\[1ex]
   50  & 1.008 & 0.950502 & $2.79336\times10^{-10}$\\[1ex]
   50  & 1.009 & 0.962798 & $1.09198\times10^{-15}$\\[1ex]
   \hline
    55  & 1.007 & 0.945514 & $2.6241\times10^{-8}$\\[1ex]
   55  & 1.008 & 0.960255 & $8.18258\times10^{-16}$\\[1ex]
   55  & 1.009 & 0.970656 & $5.80923\times10^{-20}$\\[1ex]
   \hline
    60  & 1.007 & 0.955174 & $2.28734\times10^{-14}$\\[1ex]
    60& 1.008 & 0.967833 & $2.40434\times10^{-20}$\\[1ex]
   60& 1.009 & 0.976722 & $1.97966\times10^{-23}$\\[1ex]
   \hline
   
   \end{tabular}

\end{table}

\begin{figure}[hbt!]
\begin{center}
\includegraphics[height=2.5in]{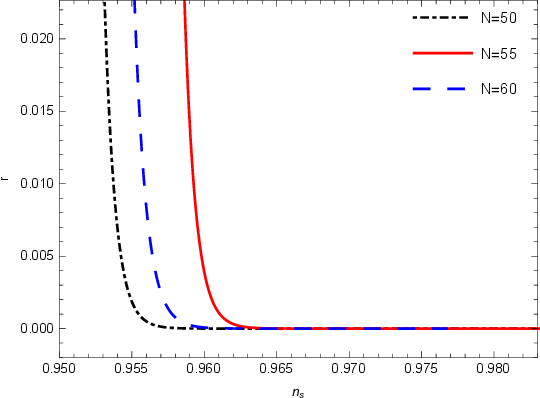}
\caption{The figure shows the tensor-to-scalar ratio $r$ with respect to the scalar spectral index $n_{s}$ for different values of $N$ for model-1. The other paramet-ers are $m = 0.75;
c1 = 1;
\gamma = 0.99;$}
\label{figscale3}
\end{center}
\end{figure}

\begin{figure}[hbt!]
\begin{center}
\includegraphics[height=2.5in]{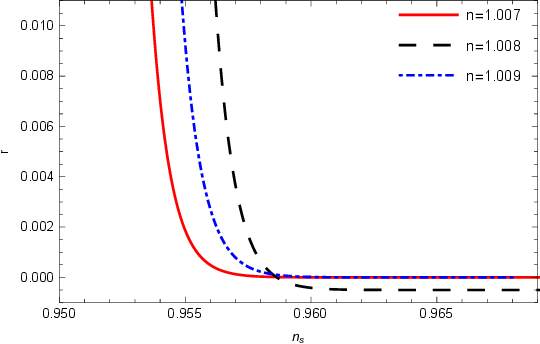}
\caption{ $r$ vs $n_{s}$ for different values of the con-stant $n$ for model-1. Other parameters are $m = 0.75,
c1 = 1,
\gamma = 0.99$.}
\label{figscale4}
\end{center}
\end{figure}

\begin{figure}[hbt!]
\begin{center}
\includegraphics[height=2.5in]{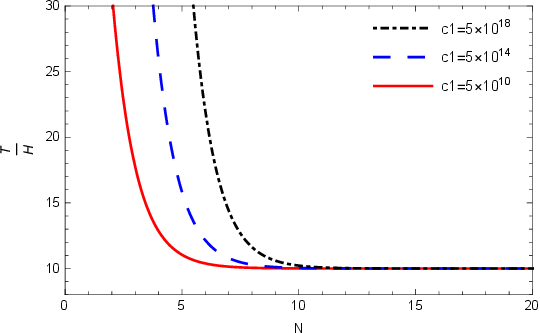}
\caption{ Plot of $T/H$ against $N$ for different val-ues of the $c_{1}$ for mod-el-1. Other parameters are $
m = 202,
n = -500.
$}
\label{figscale5}
\end{center}
\end{figure}

\begin{figure}[hbt!]
\begin{center}
\includegraphics[height=2.5in]{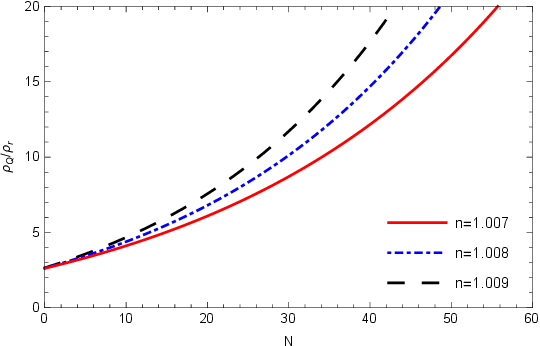}
\caption{Ratio of densities $\rho_{Q}/\rho_{r}$ against $N$ for different values of $n$ for model-1. Other parameters are $c1 = 1,
m = 4.5.$}
\label{figscale6}
\end{center}
\end{figure}

\subsubsection{Toy model-2}
We consider the second toy model as
\begin{equation} \label{u2}
f(Q)=mQ~ln(nQ)
\end{equation}
where $m$ and $n$ are constants. This functional form has deep connections to holographic principles and the AdS/CFT correspondence. In holographic cosmology, the bulk $f(Q)$ gravity theory is dual to a conformal field theory (CFT) living on the boundary. The logarithmic terms arise from the conformal anomaly of the boundary CFT. The logarithmic form also emerges from information-theoretic considerations. The holographic principle suggests that the information content of a cosmological region is encoded on its boundary. The entanglement entropy between different regions scales logarithmically with the correlation length, leading to gravitational actions with logarithmic dependence on curvature invariants. For the logarithmic model (\ref{u2}), the effective coupling is
\begin{equation}
g_{\text{eff}}^2 \approx \frac{|m|}{M_p^2} \left[ \ln(nQ) + 1 \right]^2.
\end{equation}
The microscopic derivation imposes several constraints:\\
1. \textbf{Unitarity bounds}: The couplings must satisfy $g_{\text{eff}} \lesssim 4\pi$ to maintain perturbative unitarity.\\
2. \textbf{Thermal equilibrium}: The interaction rate must exceed the Hubble rate: $\Gamma/H \gtrsim 1$ for efficient thermalization.\\
3. \textbf{Back-reaction bounds}: The energy density in produced particles must not exceed the $f(Q)$ energy density during inflation.\\
4. \textbf{Reheating consistency}: The model must smoothly connect to the radiation-dominated era with the correct thermal history.\\
These constraints provide upper and lower bounds on the model parameters $m$, $n$, and $\gamma$, making the theory predictive and testable.

Substituting the equations (\ref{u2}) and (\ref{t}) into  equation (\ref{s}), we obtain the Hubble parameter in terms of e-fold-ing number N as 
\begin{equation} 
H(N)=\frac{e^{c_{1}e^{4mN}}}{\sqrt{6n}e}
\end{equation}
Because both the slow-roll parameters depend on a function of $Q$ and $Q$ can be expressed as a Hubble parameter, both parameters are written in terms of the e-fold-ing number $N$ as 
\begin{equation}
\epsilon_{1}(N)=-4mc_{1}e^{4mN}
\end{equation}

\begin{equation}
\epsilon_{2}(N)=2m
\end{equation}
  
The warm inflation parameter $\beta$ can be written in terms of the number of e-folds as 
\begin{equation}
\beta=-\gamma m\left(2c_{1}e^{4mN}+1\right)
\end{equation}
For the high dissipative parameter, $G(\delta) \approx 0.00185~ \delta^{2.315}$ \cite{w9}.
The scalar spectral index and the tensor-to-scalar ratio for this model can be found just like with the previous model using the equations (\ref{nss1}) and (\ref{nss2}), respectively.

\begin{table}
\caption{Numerical values of the scal-ar spectral index and tensor-to-scalar ratio for different values of $N$ and the model parameter $m$ for model-2. The other con-stants are taken as  $c_1=-0.004, C_{T}=0.67, n=0.5, \gamma=0.57.$}
    \label{tab:my_label}
    \vspace{5mm}
    \centering
    \begin{tabular}{||l|c|c|r||}
    \hline
        N & m & $n_{s}$ &r\\ [1ex]
   
   \hline \hline
   50  & 0.005 & 0.979758 & 0.00337578\\[1ex]
   50  & 0.008 & 0.968444 & 0.00640455\\[1ex]
   50  & 0.011 & 0.95869 & 0.0102024\\[1ex] 
   \hline
     55  & 0.005 & 0.979825 & 0.00370985\\[1ex]
     55  & 0.008 & 0.968763 & 0.00709363\\[1ex]
     55  & 0.011 & 0.959824 & 0.0114458\\[1ex]
   \hline
    60& 0.005 & 0.979898 & 0.00404623\\[1ex]
    60& 0.008 & 0.969138 & 0.00780808\\[1ex]
    60  & 0.011 & 0.961238 & 0.0127727\\[1ex]
   \hline
   
   \end{tabular}

\end{table}


\label{figscale}

\begin{figure}[hbt!]
\begin{center}
\includegraphics[height=2.5in]{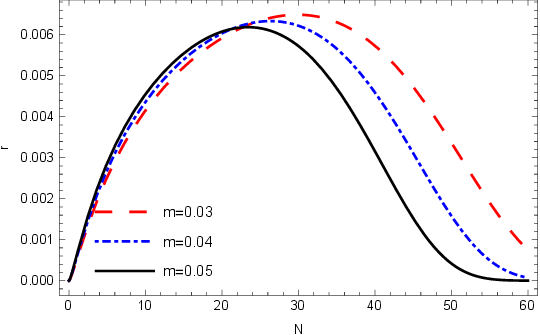}
\caption{The figure shows the tensor-to-scalar ratio $r$ against $N$ for different values of parameter $m$ for model-2. Other parameters are considered as $c_{1}=-0.87$,~ $n=0.5$,~~ $C_{T}=0.67$.}
\label{figscale7}
\end{center}
\end{figure}

\begin{figure}[hbt!]
\begin{center}
\includegraphics[height=2.5in]{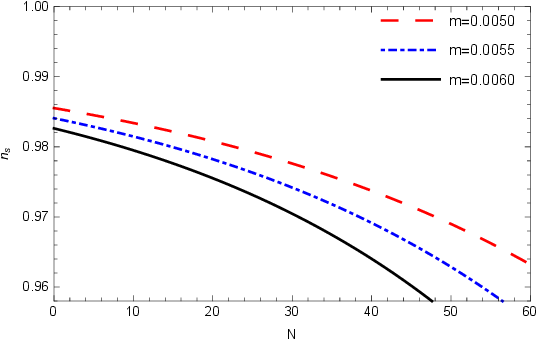}
\caption{The figure shows the Scalar spectral index $n_{s}$ for model-2 against $N$ for different values of $m$. Other parameters are $c1 = 0.37,
m = 0.006,
\gamma = -0.27.$}
\label{figscale8}
\end{center}
\end{figure}

\begin{figure}[hbt!]
\begin{center}
\includegraphics[height=2.5in]{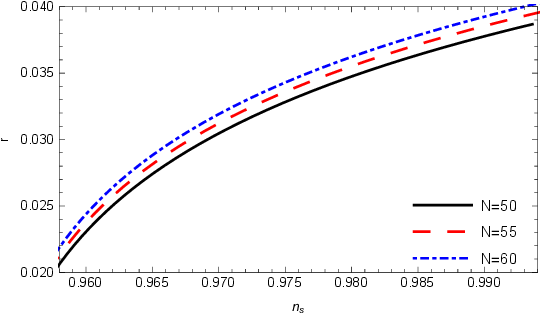}
\caption{Tensor-to-scalar ratio $r$ with respect to the scalar spectral index $n_{s}$ for different values of $N$ for mod-el-2.Other parameters are $c_1 = -0.37;
n = 0.023;
\gamma = 0.87;
$}
\label{figscale9}
\end{center}
\end{figure}

\begin{figure}[hbt!]
\begin{center}
\includegraphics[height=3.5in]{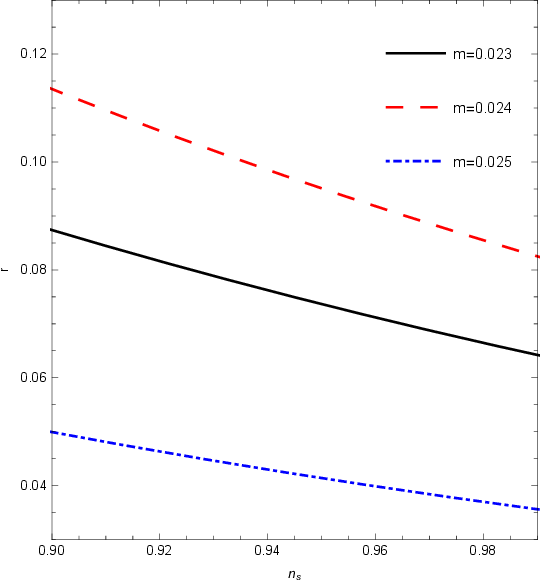}
\caption{The tensor-to-scalar ratio $r$ against the scalar spectral index $n_{s}$ for different values of $m$ for model-2. Other parameters are $c_1 = -0.87,
n = 0.5,
\gamma = 0.99.$}
\label{figscale10}
\end{center}
\end{figure}

\begin{figure}[hbt!]
\begin{center}
\includegraphics[height=2.5in]{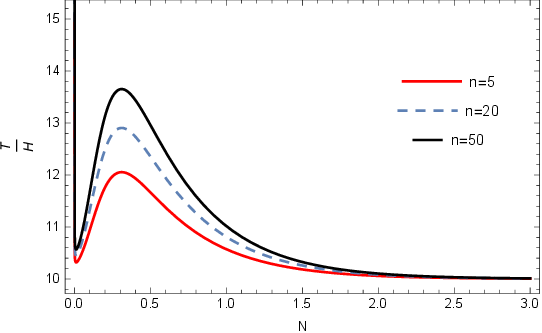}
\caption{ Plot of $T/H$ against $N$ for different values of the constant $n$ for model-2. Other parameters are $c_1 = 0.17,
n = 50,
c_{r} = 7.$}
\label{figscale11}
\end{center}
\end{figure}

\begin{figure}[hbt!]
\begin{center}
\includegraphics[height=2.5in]{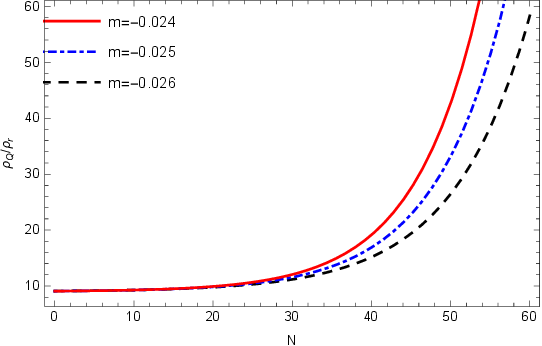}
\caption{Plot of density ratio $\rho_{Q}/\rho_{r}$ against $N$ for different values of $m$ for model-2. Other parameters are $c_1 = 200,
m = -0.030.$}
\label{figscale12}
\end{center}
\end{figure}

\section{Results}
The outcome of the study should be correlated with observational data to confirm its validity, or the observational data could be employed to restrain the model's free constants. We have expressed the inflation parameters $\epsilon_{1}$, $\epsilon_{2}$, $r$, $\beta$, and $n_{s}$ in terms of the e-folding number $N$ in the preceding section. For a warm inflationary scenario, the slow roll parameters should obey the conditions $\left|\epsilon_{1}\right|<<1$,~~$\left|\epsilon_{2}\right|<<1$, and~~$\left|\beta\right|<<1$. In Figs.(\ref{figscale1}) and (\ref{figscale2}), we have plotted the tensor-to-scalar ratio $r$ and scalar spectral index $n_{s}$ against the e-folding number $N$ for model-1. We see that in the region $0<N<60$, both the parameters stay under the value $1$, satisfying the condition for warm inflation. The upper limit of $r$ is $r<0.064$, based on the most recent observational data. We see that this condition is satisfied in our work for model-1. The plot indicates that $r$ falls within the acceptable range for our model. According to the most recent observational data, $n_s$ falls between $0.9642$ and $0.0042$. The scalar spectral index $n_{s}$ of our model approximately lies within the range ( on the slightly higher side) indicated above. In all the plots, we have considered the evolution in terms of the number of e-folds from higher $N$ to lower $N$, with $N=0$ indicating the end of inflation. In Figs.(\ref{figscale7}) and (\ref{figscale8}), similar plots have been generated for model-2. In Fig.(\ref{figscale7}), we see that $r$ reaches a maximum value and then decays, tending towards $0$. However, in the entire range of $N$, the tensor-to-scalar ra-tio remains in the range $r<0.064$, which is observation-ally acceptable. Moreover, in Fig.(\ref{figscale8}), we see that the scal-ar-spectral index lies in the acceptable range.

In Figs.(\ref{figscale3}) and (\ref{figscale4}), $r$ is plotted against $n_{s}$ for different values of e-folding number $N$ and model parameter $n$, re-spectively, for model-1. In both the plots, we see that as $r$ de-cays, $n_{s}$ almost remains stagnant. As $r$ tends to $0$, the tra-jectories nearly coincide, showing reduced dependence on the e-folding number and the model parameter $n$. Here, it is important to note that both $r$ and $n_{s}$ remain in the acceptable range for warm inflation. Here, the values of $n_{s}$ and $r$ correspond to the values at the time of horizon crossing. In the table $1$, we have provided some numerical values of $r$ and $n_{s}$ for different values of $N$ and $n$. It can be ob-served that by increasing the number of e-folds, $n_{s}$ in-creases and $r$ decreases for a given value of constant $n$. Moreover, as $n$ increases for a given value of $N$, $n_{s}$ increases, and $r$ decreases. In Figs.(\ref{figscale9}) and (\ref{figscale10}), $r$ is plotted against $n_{s}$ for different values of the e-folding number $N$ and model parameter $m$, respectively, for model-2. In Fig.(\ref{figscale9}), we see that $r$ and $n_{s}$ are positively correlated, i.e., they increase and decrease simultaneously with one another. Moreover, for greater values of $N$, we get greater values of both $r$ and $n_{s}$. In Fig.(\ref{figscale10}), it is seen that as $n_{s}$ increases, there is a corresponding decrease in the value of $r$. This shows a negative correlation between the two parameters. From table $2$, it is evident that for a given value of constant $m$, $r$ and $n_{s}$ both increase as the number of e-folds increases. Conversely, for a given value of $N$, $r$ rises and $n_{s}$ increases as $m$ increases.

Fig.(\ref{figscale5}) shows the plots of $T/H$ against the number of e-folds $N$ for different values of the constant $c_{1}$ for model-1. From the figure, it is evident that the requirement $T/H > 1$ is maintained throughout the inflationary period. This shows that temperature dominates over the Hubble ex-pansion parameter, indicating a warm inflationary scen-ario. The corresponding $T/H$ vs $N$ plot for model-2 is shown in Fig.(\ref{figscale11}). Here, different trajectories are generated for different values of the parameter $n$. We also see that the trajectories remain at the level $T/H>1$, indicat-ing a warm inflationary phase. Moreover, for higher val-ues of $n$, we get a greater dominance of $T$ over $H$, show-ing intensified warm inflation.

The ratio of energy density of $f(Q)$ gravity $\rho_{Q}$ to radiation energy density $\rho_{r}$ is plotted against $N$ in Figs.(\ref{figscale6}) and (\ref{figscale12}) for model-1 and model-2, respectively. In both the cases, it is seen that the ratio was high at the beginning of the inflation, but it dropped toward the end ($N\rightarrow0$), and the various trajectories approached one another. Initially, the value of the ratio was significantly higher than $1$, indicating that $\rho_{Q}$ was the primary driver of inflation. $\rho_{r}$ became more significant as the inflation continued and approached the end. The ratio decreased as the inflation came to an end, indicating that the two densities became comparable to each another.

\section{Discussion and Conclusion}
Similar to the classical cold inflation, warm inflation also presents a framework for exploring the dynamics of the early universe.  To address the drawbacks of the conventional cosmological model, the inflationary paradigm was presented. The matter field progressively rolls to its flat potential in cold inflation as it does not interact with radiation. In contrast, in the warm inflationary scenario, energy is transferred from the inflaton to the radiation field during slow-roll as a result of interactions between the inflaton and other fields. When inflation stops, the in-flaton completely decays into radiation, preventing the universe from going into a very cold phase. Consequently, a distinct reheating phase is not required as the universe transitions into a radiation-dominated phase.

In line with the conventional big bang model, warm inflation has emerged as a substitute theory for cold inflation. Concurrently, alternative theories of gravity have been created to address the drawbacks of General Relativity.  In this work, we have investigated warm inflation in the background of $f(Q)$ gravity considering a flat FRW spacetime. We have studied the inflationary scenario mainly in the strong dissipative regime $(\delta>> 1)$. The dissipation term in warm inflation is determined by accounting for the interactions between scalar and other fields. The field equations of $f(Q)$ gravity have been used, and we have reconstructed the Hubble parameter as a function of the e-folding number $N$. Slow roll parameters significantly influence warm inflation, thus, the consistency of the slow-roll approximation depends on a set of slow-roll parameters. Two different $f(Q)$ toy-models were considered to validate our investigation. We have demonstrated that these slow-roll parameters will meet the heated inflationary conditions for both the models considered. Plots were generated for the slow-roll parameters, and it was seen that there was a good agreement with the observational data. The values obtained show that our model is aligned with the Starobinsky-like inflationary model and is per-fectly consistent with the Planck and BICEP/Keck data.

Two conditions are taken into consideration in a warm inflationary scenario:  i) the density contribution from modified gravity predominates over radiation density, and ii) thermal fluc-tuation prevails over quantum fluctuation, i.e., $T>H$. We have thoroughly confirmed these two prerequisites in this work.  We have confirmed that the quantum fluctuation is subordinated to thermal fluctu-ation in both the toy-models of $f(Q)$ gravity. Further-more, the high value of $f(Q)$ energy density at inflation was found to have reduced as the universe evolved, due to energy being transferred to radiation. In the inflation-ary regime, it was finally observed that $\frac{\rho_{Q}}{\rho_{r}}>>1 $, confirming that the $f(Q)$ energy density is the driving force behind the inflation and can be compared to a scalar field. However, as inflation slows down, the two densities gradually coincide with one another. It should be noted that in this work, we have considered the evolution in terms of the number of e-folds from higher $N$ to lower $N$, with $N=0$ indicating the end of inflation.

Stability refers to whether small deviations in the background dynamics (field value, slow-roll parameters, initial conditions) lead to converging (stable) or diverging (unstable) evolution of inflationary observables like $n_s$ and $r$. Thus it is important to study the stability of the inflationary observables to gain greater insights into the system. This can be done by a dynamical system approach or by studying the features of the shape of the potential. As this requires a fully dedicated work, it will feature in one of our future projects. Examining our warm inflation model with $f(Q)$ dark energy in light of the Swampland conjectures might be an intriguing area for further research.  Many inflationary models have faced serious difficulties as a result of these conjectures, which offer standards for effective field theories that can be reliably included into quantum gravity.  Because of its unique dynamics and lower need for slow-roll, warm inflation has shown promise in resolving some of the issues, especially in relation to the de Sitter conjecture. Moreover, constraining model parameters in the light of observational data is an important project to be considered for future work.

\section*{Acknowledgments}
P.R. and F.R. acknowledges the Inter-University Centre for Astronomy and Astrophysics (IUCAA), Pune, India, for granting a visiting associateship. The authors thank the anonymous referee for the constructive comments that helped them to improve the quality of the manuscript. FR is also thankful to ANRF, DST for financial support.

\section*{Data Availability Statement}

No data was generated or analyzed in this study.

\section*{Conflict of Interest}

There are no conflicts of interest.

\section*{Funding Statement}

There is no funding to report for this article.

\end{document}